\documentclass{jpp}

\usepackage{bm}
\usepackage{graphicx}
\usepackage{epstopdf, epsfig}
\usepackage{hyperref}
\usepackage{xcolor}			
\usepackage{amsmath}		
\usepackage{amssymb}		
\usepackage{mathtools}
\usepackage{thmtools} 		
\usepackage{mathrsfs} 		
\usepackage{url}			
\usepackage{afterpage}		
\usepackage{subcaption}		
\usepackage{setspace}
\usepackage{multirow}
\usepackage{booktabs}
\usepackage{rotating}

\newcommand{\para}{\parallel}								
\newcommand{\diffIL}[2]{\mathrm{d}#1/\mathrm{d}#2}	
\newcommand{\diff}[2]{\frac{\mathrm{d}#1}{\mathrm{d}#2}}	
\newcommand{\pdiff}[2]{\frac{\partial#1}{\partial#2}}		
\renewcommand{\tt}{\tilde{\theta}}							
\renewcommand{\d}{\textrm{d}}								
\newcommand{\kperp}{k_{\perp}}								
\newcommand{\kpara}{k_{\para}}								
\newcommand{\vperp}{v_{\perp}}								
\newcommand{\wperp}{w_{\perp}}

\newcommand{\zhat}{\hat{\boldsymbol{z}}}						
\renewcommand{\b}[1]{\boldsymbol{#1}}							
\newcommand{\bx}{\boldsymbol{x}}								
\newcommand{\bv}{\boldsymbol{v}}								
\newcommand{\bE}{\boldsymbol{E}}								
\newcommand{\bB}{\boldsymbol{B}}								
\newcommand{\grad}{\nabla}									
\newcommand{\curl}{\nabla\times}							
\newcommand{\va}{v_{\rm A}}										

\shorttitle{Stochastic proton heating by KAW Turbulence}
\shortauthor{I. W. Hoppock et al.}

\title{Stochastic proton heating by kinetic-Alfv\'en-wave turbulence in moderately high-$\beta$ plasmas}

\author{
Ian W. Hoppock\aff{1}
\corresp{\email{ian.hoppock@unh.edu}},
Benjamin D. G. Chandran\aff{1},\\
Kristopher G. Klein\aff{2,3}, 
Alfred Mallet\aff{1,4}, and
Daniel Verscharen\aff{1,5}
}

\affiliation{
\aff{1}Space Science Center, University of New Hampshire, Durham, NH, 03824, USA
\aff{2}Lunar and Planetary Laboratory, University of Arizona, Tucson, AZ, 85719, USA
\aff{3}CLASP, University of Michigan, Ann Arbor, MI, 48109, USA
\aff{4}Space Sciences Laboratory, University of California, Berkeley, CA, 94720, USA
\aff{5}Mullard Space Science Laboratory, University College London, Dorking, RH5 6NT, UK 
}

\begin{document}

\maketitle

\begin{abstract}
Stochastic heating refers to an increase in the average 
magnetic moment of charged particles interacting with electromagnetic fluctuations whose frequencies are smaller than the particles' cyclotron frequencies. This type of heating arises when the amplitude of the gyroscale fluctuations exceeds a certain threshold, causing particle orbits in the plane perpendicular to the magnetic field to become stochastic rather than nearly periodic.
We consider the stochastic heating of protons by Alfv\'en-wave (AW) and kinetic-Alfv\'en-wave (KAW) turbulence, which may make an important contribution to the heating of the solar wind. 
Using phenomenological arguments, we derive the
stochastic-proton-heating rate in plasmas in which $\beta_{\rm p} \sim
1-30$, where $\beta_{\rm p}$ is the ratio of the proton pressure to
the magnetic pressure. (We do not consider the $\beta_{\rm p} \gtrsim
30$ regime, in which KAWs at the proton gyroscale become
non-propagating.) We test our formula for the stochastic-heating rate
by numerically tracking test-particle protons interacting with a
spectrum of randomly phased AWs and KAWs.  Previous studies have
demonstrated that at $\beta_{\rm p} \lesssim 1$, particles are
energized primarily by time variations in the electrostatic potential
and thermal-proton gyro-orbits are stochasticized primarily by
gyroscale fluctuations in the electrostatic potential. In contrast, at
$\beta_{\rm p} \gtrsim 1$, particles are energized primarily by the
solenoidal component of the electric field and thermal-proton
gyro-orbits are stochasticized primarily by gyroscale fluctuations in
the magnetic field.  \end{abstract}

\section{Introduction}\label{sec:intro}

In the mid-twentieth century several authors published hydrodynamic models of the solar wind that imposed a fixed temperature at the coronal base and took thermal conduction to be the only heating mechanism \citep[e.g.,][]{parker1958,parker1965,hartelsturrock1968,durney1972}. These models were unable to explain the high proton temperatures and fast-solar-wind speeds observed at a heliocentric distance~$r$ of 1 astronomical unit (au) for realistic values of the coronal temperature and density, indicating that the fast solar wind is heated primarily by some mechanism other than thermal conduction.
\citet{parker1965} and \citet{coleman1968} proposed that Alfv\'en waves (AWs) and AW turbulence provide this additional heating.
Support for this suggestion can be found in the many spacecraft observations of AW-like turbulence in the solar wind \citep[see][]{belcher71,tu1995,bale2005}, remote observations of AW-like fluctuations in the solar corona \citep[see][]{tomczyk2007,depontieu2007}, and the agreement between AW-driven solar-wind models and solar-wind temperature, density, and flow-speed profiles \citep{cranmer07,verdini10,chandran11,vanderholst14}.

AWs oscillate at a frequency $\omega = \kpara \va$, where $\kpara$
($\kperp$) is the component of the wave vector $\b{k}$ parallel
(perpendicular) to the background magnetic field, $\bm{B}_0$,
$\va = B_0 / \sqrt{4\pi n_{\rm p}m}$ is the Alfv\'en speed,
$n_{\rm p}$ is the proton number density, and $m$ is the proton
mass.\footnote{We neglect the mass density of electrons and heavy
  ions.}  In AW turbulence, interactions between counter-propagating
AWs cause AW energy to cascade from larger to smaller scales.  This
energy cascade is anisotropic, in the sense that the small-scale AW
``eddies,'' or wave packets, generated by the cascade vary much more
rapidly perpendicular to the magnetic field than
along the magnetic field \citep[e.g.,][]{shebalin1983,gs95,cho00,horbury08,podesta13,chen16}. As a
consequence, within the inertial range (scales larger than the
thermal-proton gyroradius~$\rho_{\rm th}$ and smaller than the outer
scale or driving scale), $\omega \ll \Omega$, where $\Omega$ is the
proton cyclotron frequency.  At $k_\perp \rho_{\rm th} \sim 1$, the AW
cascade transitions to a kinetic-Alfv\'en-wave (KAW) cascade
\citep{sch2009}.

Studies of the dissipation of low-frequency ($\omega \ll \Omega$),
anisotropic, AW/KAW turbulence based on linear wave damping
\citep[e.g.,][]{quataert1998,2howes2008} conclude that AW/KAW
turbulence leads mostly to parallel heating of the particles (i.e.,
heating that increases the speed of the thermal motions along
$\bm{B}$). On the other hand, perpendicular ion heating is the
dominant form of heating in the near-Sun solar wind
\citep{esser1999,marsch06,cranmer2009b,hellinger2013}. This discrepancy
suggests that AW/KAW turbulence in the solar wind dissipates
via some nonlinear mechanism
\citep[e.g.][]{dmitruk2004,markovskii2006,letal2009,sch2009,cetal2010,servidio2011b,lynn2012,xetal2013,kawazura18}. 
This suggestion is supported by studies that find a correlation between ion temperatures and fluctuation amplitudes in solar-wind measurements and numerical simulations \citep[e.g.,][]{wu13b,hughes17a,groselj17,vech18b}.

In this paper, we consider one such nonlinear mechanism: stochastic heating. In stochastic proton heating, AW/KAW fluctuations at the proton gyroscale have sufficiently large amplitudes that they disrupt the normally smooth cyclotron motion of the protons, leading to non-conservation of the first adiabatic invariant, the magnetic moment \citep{mcchesney1987,jc2001,chen2001,chaston2004,fiksel2009,xetal2013}. \citet{cetal2010} used phenomenological arguments to derive an analytical formula for the stochastic-heating rate at $\beta_{\rm p} \lesssim 1$, where 
$\beta_{\rm p}$ is the ratio of the proton pressure to the magnetic pressure (see~(\ref{eq:defbetap})). In Section~\ref{sec:sh} we use phenomenological arguments to obtain an analytic formula for the proton-stochastic-heating rate in low-frequency AW/KAW turbulence when $\beta_{\rm p} \sim 1-30$. We limit our analysis to $\beta_{\rm p} \lesssim 30$, since KAWs become non-propagating at $k_\perp \rho_{\rm th}=1$ at larger~$\beta_{\rm p}$ values (see Appendix~\ref{sec:limitations} and \cite{hellinger00,kawazura18,kunz18}).
In Section~\ref{sec:test} we present results from simulations of test particles interacting with a spectrum of randomly phased AWs/KAWs, which we use to test our analytic formula for the stochastic-heating rate.
 Throughout this paper, we focus on perpendicular proton heating
  rather than parallel proton heating. Stochastic heating can in
  principle augment the parallel proton heating that results from linear damping of AW/KAW turbulence at $\beta_{\parallel \rm p} \gtrsim 1$, but we leave a discussion of this possibility to future work.

\section{Stochastic proton heating by AW/KAW turbulence at the proton gyroscale}
\label{sec:sh} 

A proton interacting with a uniform background magnetic field~$\bm{B}_0$ and fluctuating electric and magnetic fields~$\delta \bm{E}$ and~$\delta \bm{B}$ undergoes nearly periodic motion in the plane perpendicular to~$\bm{B}_0$ if $\delta E$ and $\delta B$ are sufficiently small or $L/\rho$ is sufficiently large, where
$L$ is the characteristic length scale of $\delta \bm{E}$ and $\delta \bm{B}$, $\rho = \vperp / \Omega$ is the proton's gyroradius, $\bm{v}_\perp$ is the component of the proton's velocity~$\bm{v}$ perpendicular to the magnetic field, $\Omega = qB_0/mc$ is the proton gyrofrequency, $m$ and $q$ are the proton mass and charge, and $c$ is the speed of light. When (1) the proton's motion in the plane perpendicular to~$\bm{B}_0$ is nearly periodic and (2) $\Omega \tau \gg 1$, where $\tau$ is the characteristic time scale of~$\delta \bm{E}$ and $\delta \bm{B}$,  the proton's magnetic moment $\mu = m\vperp^2/2B_0$ is almost exactly conserved \citep{kruskal}.

Perpendicular heating of protons (by which we mean a secular increase in the average value of~$\mu$) requires that one of the above two conditions for $\mu$ conservation be violated.
For example, Alfv\'en/ion-cyclotron waves can cause perpendicular proton heating via a cyclotron resonance if $\Omega \tau \sim 1$ 
\citep{hollweg02}. Alternatively, low-frequency AW/KAW fluctuations can cause perpendicular proton heating if their amplitudes at $k_\perp \rho \sim 1$ are sufficiently large that the proton motion in the plane perpendicular to~$\bm{B}_0$ becomes disordered or ``stochastic''
\citep{mcchesney1987,johnson01,chen01,chaston04,fiksel09}.

We focus on this second type of heating, stochastic heating, and on ``thermal'' protons, for which 
\begin{equation}
v_\parallel \sim w_\parallel \qquad  \vperp \sim \wperp \qquad \rho \sim \rho_{\rm th},
\label{eq:thermal} 
\end{equation} 
where $\wperp = \sqrt{2k_{\rm B} T_{\perp \rm p}/m}$ and $w_\parallel = \sqrt{2k_{\rm B} T_{\parallel \rm p}/m}$ are the  perpendicular
and parallel thermal speeds,  $T_{\perp \rm p}$ and $T_{\parallel \rm p}$ are the perpendicular and parallel proton temperatures,  $k_{\rm B}$ is Boltzmann's constant,
and $\rho_{\rm th } = \wperp/\Omega$ is the thermal-proton gyroradius. 
We restrict our attention to the contribution to the stochastic-heating rate from turbulent AW/KAW fluctuations with 
\begin{equation}
\lambda \sim \rho_{\rm th} \qquad k_\perp \rho_{\rm th} \sim 1 ,
\label{eq:lambdarho} 
\end{equation} 
where $\lambda$ is the length scale of the fluctuations measured perpendicular to the background magnetic field, and to 
\begin{equation} 
\beta_{\rm p} \equiv \frac{8 \pi n k_{\rm B} T_{\rm p}}{B_0^2}  \sim 1 - 30,
\label{eq:defbetap} 
\end{equation} 
where 
\begin{equation}
T_{\rm p} = \frac{2T_{\perp \rm p} + T_{\parallel \rm p}}{3}.
\label{eq:defTp} 
\end{equation} 
As mentioned above and discussed further in Appendix~\ref{sec:limitations}, KAWs at $k_\perp \rho_{\rm th}  = 1$ become non-propagating at significantly larger values of~$\beta_{\rm p}$ \citep[see also][]{hellinger00,kawazura18,kunz18}.
For simplicity, we assume that
\begin{equation}
T_{\perp \rm p} \sim T_{\parallel \rm p} \qquad w_\perp \sim w_\parallel,
\label{eq:iso2} 
\end{equation} 
which implies that 
\begin{equation}
\beta_{\rm p} \sim \frac{\wperp^2}{v_{\rm A}^2} \sim \frac{w_\parallel^2}{v_{\rm A}^2},
\label{eq:betaperp} 
\end{equation} 
and that $T_{\rm e} \sim T_{\rm p}$, where $T_{\rm e}$ is the electron temperature.
We also assume that
\begin{equation}
\delta B_{\rho} \ll B_0,
\label{eq:smalldB}  
\end{equation} 
where $\delta B_{\rho}$ is the rms amplitude of the magnetic fluctuations with $\lambda \sim \rho_{\rm th}$, and that the fluctuations are in critical balance \citep{gs95}, which implies that
\begin{equation}
\frac{\delta v_\rho }{\rho_{\rm th}} \sim \frac{v_{\rm A}}{l},
\label{eq:CB} 
\end{equation} 
where $l$ is the correlation length of the gyroscale AW/KAW fluctuations measured parallel to the background magnetic field, and
 $\delta v_\rho$ is the rms amplitude of the $\bm{E}\times \bm{B}$ velocity of the AW/KAW fluctuations with $\lambda \sim \rho_{\rm th}$.
Since the linear and nonlinear time scales are comparable in the critical-balance model, we take the ratios of the amplitudes of different fluctuating variables to be comparable to the ratios that arise for linear AW/KAWs at $k_\perp \rho_{\rm th} \sim 1$, which, given 
(\ref{eq:lambdarho}) and (\ref{eq:defbetap}), implies that
\begin{equation}
\delta B_{\parallel \rho} \sim \delta B_{\perp \rho} \sim \delta B_\rho
\qquad \frac{\delta B_\rho}{B_0} \sim \frac{\delta v_\rho}{v_{\rm A}},
\label{eq:eigen} 
\end{equation} 
where $\delta B_{\parallel \rho}$ and $\delta B_{\perp \rho}$ are, respectively, the rms amplitudes of the components of the fluctuating magnetic field parallel and perpendicular to~$\bm{B}_0$ \citep{tenbarge12}. Equations~(\ref{eq:lambdarho}) through~(\ref{eq:eigen}) imply that
\begin{equation}
\omega \sim \frac{v_{\rm A}}{l} \sim \frac{\delta v_{\rho}}{\rho_{\rm th}} \sim \Omega \frac{\delta v_\rho}{\wperp} \sim \Omega \beta_{\rm p}^{-1/2} \frac{\delta v_\rho}{v_{\rm A}}
\sim \Omega \beta_{\rm p}^{-1/2} \frac{\delta B_\rho}{B_0}
\ll \Omega.
\label{eq:omegasmall} 
\end{equation} 

\subsection{Stochastic motion perpendicular to the magnetic field}
\label{sec:perp} 

To understand how gyroscale AW/KAW fluctuations modify a proton's motion, we cannot use the adiabatic approximation \citep{nor63}, which assumes $\lambda \gg \rho$. Nevertheless, we can still define an effective guiding center
\begin{equation}
\bm{R} = \bm{r} + \frac{\bm{v} \times \bm{\hat{b}}}{\Omega},
\label{eq:defR} 
\end{equation} 
where $ \bm{\hat{b}} = \bm{B}/B$. This effective guiding center 
is always a distance~$\rho$ from the particle's position~$\bm{r}$ and is, at any given time,
the location about which the particle attempts to gyrate under the influence of the Lorentz force. We find it useful to focus on $\bm{R}$ rather than~$\bm{r}$ because the motion of~$\bm{R}$ largely excludes the high-frequency cyclotron motion of the proton. Upon taking the time derivative of~(\ref{eq:defR}) and making use of the relations
$ \diffIL{\bm{r}}{t} = \bm{v}$ and $ \diffIL{\bm{v}}{t}= (q/m)(\bm{E} + \bm{v} \times \bm{B}/c)$, we obtain
\begin{equation}
\diff{\bm{R}}{t}  = v_\parallel \bm{\hat{b}} + \frac{c \bm{E} \times \bm{B}}{B^2}
- \frac{\bm{v} \times \bm{\hat{b}}}{\Omega} \, \frac{1}{B} \diff{B}{t}
+ \frac{\bm{v}}{\Omega} \times \diff{\bm{\hat{b}}}{t},
\label{eq:dRdt1} 
\end{equation} 
where $v_\parallel = \bm{v} \bcdot \bm{\hat{b}} $.
The perpendicular component of $\diffIL{\bm{R}}{t}$,
\begin{equation}
\left(\diff{\bm{R}}{t}\right)_\perp =  \diff{\bm{R}}{t} - \bm{\hat{b}} \left( \bm{\hat{b}} \bcdot \diff{\bm{R}}{t}\right)=\left(\bm{\hat{b}}\times 
\diff{\bm{R}}{t} \right) \times \bm{\hat{b}},
\label{eq:dRdtperp} 
\end{equation} 
can be found
by substituting the right-hand side of~(\ref{eq:dRdt1}) into the right-hand side of~(\ref{eq:dRdtperp}), which yields 
\begin{equation}
\left(\diff{\bm{R}}{t}\right)_\perp= 
\frac{c \bm{E} \times \bm{B}}{B^2}
- \frac{\bm{v} \times \bm{\hat{b}}}{\Omega} \, \frac{1}{B}\diff{B}{t}
+ \frac{v_\parallel }{\Omega} \bm{\hat{b}}\times \diff{\bm{\hat{b}}}{t}.
\label{eq:dRdtperp2} 
\end{equation}

We now estimate each term on the right-hand side of~(\ref{eq:dRdtperp2}).
Since we are considering only gyroscale fluctuations\footnote{AW fluctuations at $\lambda \gg \rho_{\rm th}$
advect both the gyroscale AW/KAW eddies and the particles at the $\bm{E}\times\bm{B}$ velocity of the large-scale AW fluctuations.}, we take the first term on the right-hand side of~(\ref{eq:dRdtperp2}) 
to satisfy the relation
\begin{equation}
\left| \frac{c \bm{E} \times \bm{B}}{B^2}\right| \sim \delta v_\rho.
\label{eq:ExB} 
\end{equation} 
To estimate the second and third terms on the right-hand side of~(\ref{eq:dRdtperp2}), we take
\begin{equation}
v_\perp \sim |v_\parallel| \sim \wperp \sim w_\parallel,
\label{eq:iso} 
\end{equation} 
which is satisfied by the majority of particles.
The time derivative of the field strength along the particle's trajectory is
\begin{equation}
\diff{B}{t} = \frac{\partial B}{\partial t} + \bm{v}_\perp \bcdot \nabla B + v_\parallel 
\bm{\hat{b}}\bcdot \nabla B.
\label{eq:dBdt0} 
\end{equation} 
As outlined above, our assumption of critical balance implies that $\lambda \ll l$ and $\omega \ll v_\perp/\rho \sim w_\perp/\rho_{\rm th}$. The second term on the right-hand side of~(\ref{eq:dBdt0}) is thus much larger than either the first or third terms,
and
\begin{equation}
\diff{B}{t} \sim \frac{\wperp \delta B_{\parallel \rho}}{\rho_{\rm th}}.
\label{eq:dBdt3} 
\end{equation} 
The second term on the right-hand side of~(\ref{eq:dRdtperp2}) thus satisfies
\begin{equation}
\left| \frac{\bm{v} \times \bm{\hat{b}}}{\Omega} \frac{1}{B} \diff{B}{t} \right| 
\sim \frac{\rho_{\rm th}}{B}\diff{B}{t}  \sim \frac{\wperp \delta B_{\parallel \rho}}{B_0} ,
\label{eq:2ndterm} 
\end{equation} 
which is larger than the first term on the right-hand side of~(\ref{eq:dRdtperp2}) by a factor of $\sim \beta_{\rm p}^{1/2}$, given~(\ref{eq:betaperp}) and (\ref{eq:eigen}).

For the moment, we assume that the second term on the right-hand side of~(\ref{eq:dRdtperp2}) is the dominant term; we discuss the third term in more detail below. If the second term is dominant, then \begin{equation}
\left|\left(\diff{\bm{R}}{t}\right)_\perp \right| \sim \frac{\wperp \delta B_{\parallel \rho}}{B_0}.
\label{eq:dRdtinst} 
\end{equation} 
During a single cyclotron period $2\pi/\Omega$, a proton passes
through an order-unity number of uncorrelated gyroscale AW/KAW eddies,
and the values of $(\d \bm{R}/\d t)_\perp$ within different gyroscale
eddies are uncorrelated. If $(\d \bm{R}/\d t)_\perp$ is small compared
to~$w_\perp$, then a proton undergoes nearly circular
gyromotion. However, if $|(\d \bm{R}/\d t)_\perp|$ is a significant fraction of~$w_\perp$, then a proton and its guiding center will move in an essentially unpredictable way, and the proton's orbit will become stochastic rather than quasi-periodic. Given~(\ref{eq:eigen}),
$|(\d \bm{R}/\d t)_\perp|$ is a significant fraction of~$w_\perp$ if the stochasticity parameter
\begin{equation}
\delta \equiv \frac{\delta B_\rho}{B_0}
\label{eq:defdelta} 
\end{equation} 
is a significant fraction of~unity.

\begin{figure}
\centerline{\includegraphics[width=\textwidth]{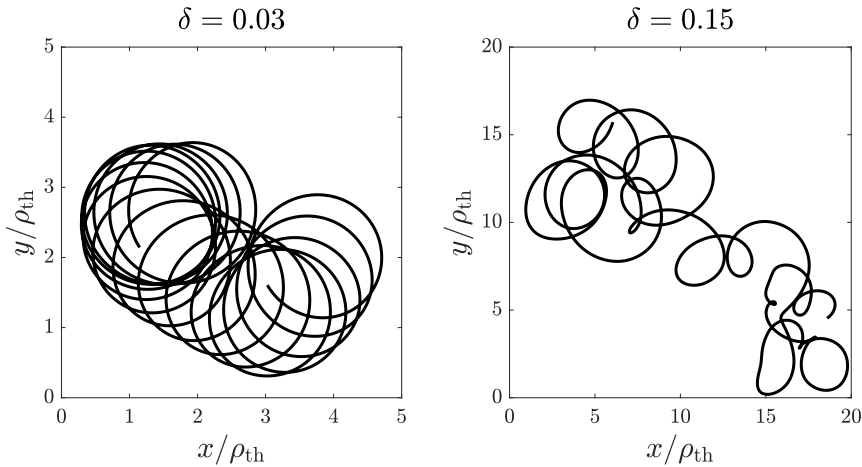}}  
\caption{Trajectories of test-particle protons interacting with a spectrum of randomly phased AWs and KAWs for different values of the stochasticity parameter~$\delta$ defined in~(\ref{eq:defdelta}).
}
\label{fig:trajectories}
\end{figure}

We illustrate how the value of~$\delta $ affects a proton's motion in figure~\ref{fig:trajectories}. We compute the particle trajectories shown in this figure by numerically integrating the equations of motion for protons interacting with randomly phased AWs and KAWs. We present the details of our numerical method and more extensive numerical results in Section~\ref{sec:test}. 
In the numerical calculation shown in the left panel of figure~\ref{fig:trajectories}, $\delta = 0.03$, and the proton's motion in the plane perpendicular to~$\bm{B}_0$ is quasi-periodic. In the numerical calculation shown in the right  panel of figure~\ref{fig:trajectories},  $\delta = 0.15$, and the proton trajectory is more
disordered or random.

We now consider the third term on the right-hand side of~(\ref{eq:dRdtperp2}). 
The instantaneous value of this term is comparable to the instantaneous value of the second term
given~(\ref{eq:eigen}) and (\ref{eq:iso}), but the third term is less effective at causing guiding-center displacements over time for the following reason.
Because of~(\ref{eq:smalldB}), the time $t_\parallel$ required for $v_\parallel$ to change by a factor of order unity is $\gg \Omega^{-1}$. 
If we integrate the third term on the right-hand side of~(\ref{eq:dRdtperp2}) from $t=0$ to $t=t_{\rm f}$, where $\Omega^{-1} \ll t_{\rm f} \ll t_\parallel$, we can treat $v_\parallel$ as approximately constant in~(\ref{eq:dRdtperp2}), obtaining
\begin{equation}
\int_0^{t_{\rm f}} \frac{v_\parallel}{\Omega} \bm{\hat{b}} \times \diff{\bm{\hat{b}}}{t} \mathrm{d}t
= \frac{v_\parallel}{\Omega_0} \frac{\bm{B}_0}{B_0} \times \Delta \bm{\hat{b}}
\label{eq:est1} 
\end{equation} 
to leading order in $\delta B_\rho/B_0$, where $\Omega_0 = qB_0/mc$ and $\Delta \bm{\hat{b}} = \bm{\hat{b}}(t_{\rm f}) - \bm{\hat{b}}(0)$ is the change in $\bm{\hat{b}}$. There is, however, no secular change in the value of $\bm{\hat{b}}$ at the proton's location; the magnetic-field unit vector merely undergoes small-amplitude fluctuations about the direction of the background magnetic field. Thus, over time, the guiding-center displacements caused by the third term on the right-hand side of~(\ref{eq:dRdtperp2}) are largely reversible and tend to cancel out. The third term is thus less effective than the second term at making proton orbits stochastic. 

When the stochasticity parameter~$\delta$ defined in
(\ref{eq:defdelta}) exceeds some threshold, the motion of a thermal proton's guiding center in the plane perpendicular to~$\bm{B}_0$ is reasonably approximated by a random walk. To estimate the time step of this random walk, we begin by defining
the cyclotron average of $(\diffIL{\bm{R}}{t})_\perp$,
\begin{equation}
\bm{v}_R(t) \equiv   \frac{\Omega}{2\pi}\int_{t - \pi/\Omega}^{t+\pi/\Omega}
 \left(\diff{\bm{R}}{t_1}\right)_\perp \mathrm{d}t_1.
\label{eq:dRdtbar} 
\end{equation} 
As stated above, during a single cyclotron period a proton's motion projected onto the plane perpendicular to~$\bm{B}_0$ carries the proton through
an order-unity number of uncorrelated gyroscale AW/KAW ``eddies.'' For simplicity, we take
the amplitude and direction of
each vector term on the right-hand side of~(\ref{eq:dRdtperp2}) to be approximately constant within any single gyroscale eddy and the values of these vector terms within different eddies to be uncorrelated.  This makes~$\bm{v}_R$ approximately equal to the average of some order-unity number of uncorrelated vectors of comparable magnitude. The amplitude of this average is comparable to the
 instantaneous value of $|(\diffIL{\bm{R}}{t})_\perp|$. 
Thus, given (\ref{eq:betaperp}), (\ref{eq:eigen}), and~(\ref{eq:dRdtinst}),
\begin{equation}
v_R \sim \frac{w_\perp \delta B_{\parallel\rho}}{B_0} \sim \beta_{\rm p}^{1/2} \delta v_\rho.
\label{eq:vR2} 
\end{equation} 

Because we are considering the effects of just the gyroscale AW/KAW eddies, $\bm{v}_R$ decorrelates 
after the proton's guiding center has moved a distance~$\sim \rho_{\rm th}$ in the plane perpendicular to~$\bm{B}_0$, which takes a time 
\begin{equation}
\Delta t \sim \frac{\rho_{\rm th}}{v_R}.
\label{eq:Dt} 
\end{equation} 
Thermal protons thus undergo spatial diffusion in the plane perpendicular to $\bm{B}_0$ with a spatial diffusion coefficient
\begin{equation}
D_\perp \sim \frac{\rho_{\rm th}^2}{\Delta t} \sim \beta_{\rm p}^{1/2} \delta v_\rho \rho_{\rm th}.
\label{eq:Dperp} 
\end{equation} 
Given~(\ref{eq:CB}), (\ref{eq:iso}), and~(\ref{eq:vR2}),
\begin{equation}
\Delta t \sim \frac{\rho_{\rm th}}{\beta_{\rm p}^{1/2}\delta v_\rho} \sim  \frac{l}{\beta_{\rm p}^{1/2} v_{\rm A}} \sim \frac{l}{v_\parallel}.
\label{eq:Dt3} 
\end{equation} 
The time required for a particle to wander a distance~$\sim \rho_{\rm th}$ perpendicular to the background magnetic field is thus comparable to the time required for the particle to traverse the parallel dimension of a gyroscale AW/KAW eddy.

\subsection{Energy diffusion and heating}
\label{sec:heating}

The total energy of a proton is given by its Hamiltonian,
\begin{equation}
H = q \Phi + \frac{1}{2m} \left( \b{p} - \frac{q}{c} \b{A} \right)^2,
\end{equation}
where $\Phi$ is the electrostatic potential, $\b{p}$ is the canonical momentum, and $\b{A}$ is the vector potential.
From Hamilton's equations,
\begin{equation} \label{derivative}
\diff{H}{t} = q\pdiff{\Phi}{t} - \frac{q\b{v}}{c} \bcdot \pdiff{\b{A}}{t},
\end{equation}
where $\bv = m^{-1}(\b{p} -q\b{A}/c)$ is the velocity, and the electric field is $\bE = -\nabla \Phi - c^{-1} \partial \b{A}/\partial t$. 
The second term on the right hand side of~\eqref{derivative}
is $q\bv\bcdot\bE_{\rm s}$, where $\bE_{\rm s} = -c^{-1} \partial\b{A}/\partial t$ is the solenoidal component of the electric field. Equation~(46) of \citet{hollweg1} gives the ratio of $E_{\rm s}$ to the magnitude of the irrotational component of the electric field~$|\nabla\Phi|$ 
for AWs/KAWs with $\kperp\rho_{\rm th}\lesssim 1$,\footnote{$\bm{E}_{\rm s}$ is nearly perpendicular to $\bm{B}_0$, as illustrated in figure~3 of \citet{hollweg1}.}
\begin{equation}
\frac{E_{\rm s}}{|\grad \Phi|} \sim  \frac{\beta_{\rm p}\omega}{\Omega}.
\end{equation}
In their treatment of stochastic heating at $\beta_{\rm p}\lesssim 1$, \citet{cetal2010} neglected the second term on the right-hand side of \eqref{derivative}, because
this term makes a small contribution to the heating rate when $\beta_{\rm p}$ is small.
Here we focus on the effects of $\bE_s$ and make the approximation that
\begin{equation}
\diff{H}{t} \sim q \bv \bcdot \bE_{\rm s}.
\label{eq:dHdt} 
\end{equation}
We show in Appendix~\ref{ap:potential} 
that the irrotational part of the electric field contributes less to the heating rate than does the solenoidal part when $\beta_{\rm p} \gtrsim 1$.  

As a proton undergoes spatial diffusion in the plane perpendicular to the background magnetic field, the electromagnetic field at its location resulting from gyroscale AW/KAW fluctuations decorrelates on the time scale~$\Delta t$ given in~(\ref{eq:Dt3}). Within each time interval of length~$\sim \Delta t$, the proton energy changes by an amount $\delta H$ (which can be positive or negative), and the values of~$\delta H$ are uncorrelated within successive time intervals of length~$\Delta t$. As a consequence, the proton undergoes energy diffusion.

To estimate the rms value of~$\delta H$, which we denote
$\Delta H$, we adopt a simple model of a proton's motion, in which the
proton's complicated trajectory is replaced by a repeating two-step
process. In the first step, the proton undergoes circular cyclotron
motion in the plane perpendicular to~$\bm{B}_0$ for a time~$\Delta
t$. In the second step, the proton is instantly translated a
distance~$\rho_{\rm th}$ in some random direction perpendicular
to~$\bm{B}_0$.\footnote{For simplicity, our model of proton motion neglects
motion parallel to~$\bm{B}$. 
This approximation is to some extent justified by~(\ref{eq:Dt3}), which implies that
a proton is unable to escape from an eddy of length~$l$ by motion
along the magnetic field in a time shorter than~$\Delta t$.
However, we return to the issue of parallel motion in Section~\ref{sec:parallel}.}

In this simple model, a proton undergoes $N \sim \Omega \Delta t \sim (v_\perp/\rho_{\rm th}) \times (l/v_\parallel) \sim l/\rho_{\rm th} \gg 1$ circular gyrations
in the plane perpendicular to~$\bm{B}_0$  during a time~$\Delta t$. Integrating~(\ref{eq:dHdt}) for a time~$\Delta t$, we obtain
\begin{equation} 
\delta H \sim q \int_0^{\Delta t}\bv(t) \bcdot \bE_{\rm s}(\bm{r}(t),t) \d t ,
\label{eq:1} 
\end{equation} 
where $\bm{r}(t)$ is the proton's position at time~$t$.
Since $\Delta t \sim \beta_{\rm p}^{-1/2} \rho_{\rm th}/\delta v_\rho$, when $\beta_{\rm p} \gtrsim 1$, the time~$\Delta t$ for a particle to diffuse across one set of gyroscale eddies is shorter than or comparable to the linear or nonlinear time scale $\rho_{\rm th}/\delta v_\rho$ of those eddies. We thus approximate the right-hand side of~(\ref{eq:1}) by setting $\bm{E}_{\rm s}(\bm{r}(t),t) = \bm{E}_{\rm s}(\bm{r}(t), 0)$ and rewrite~(\ref{eq:1}) as
\begin{equation}
\delta H 
\sim  q N \oint \bE_{\rm s}(\bm{r},0) \bcdot \d \b{l} 
\sim  q N \int_S \curl\bE(\bm{r},0) \bcdot \d \b{S} \label{3}
\sim -\frac{qN}{c} \int_S \pdiff{\bm{B}}{t}(\bm{r},0)  \bcdot \d \bm{S},
\end{equation} 
where the line integral is along the proton's path during one complete circular gyration in the plane perpendicular to~$\bm{B}_0$, the surface integral is over the circular surface~$S$ of radius~$\rho_{\rm th}$ enclosed by the gyration,
and we have used Faraday's Law $\nabla \times \bm{E} = (-1/c) \partial \bm{B}/\partial t$. The surface~$S$ is perpendicular to~$\bm{B}_0$, and $\d\b{S}$ is anti-parallel to~$\bm{B}_0$ (anti-parallel rather than parallel since $q>0$).
The rms value of $\delta H$ thus satisfies the order-of-magnitude relation
\begin{equation}
\Delta H 
\sim \frac{qN}{c} \omega_{\rm eff} \delta B_{\parallel \rho} \rho_{\rm th}^2,
\label{eq:dH1} 
\end{equation} 
where 
\begin{equation}
\omega_{\rm eff} \equiv \frac{\delta v_\rho}{\rho_{\rm th}}
\label{eq:omegaeff} 
\end{equation} 
is the nonlinear frequency of the gyroscale fluctuations.
Upon setting $q/c = \Omega m/B$, $N = \Omega \Delta t$, and $\rho_{\rm th}^2 = w_\perp^2/\Omega^2$ in~(\ref{eq:dH1}), we obtain
\begin{equation}
\Delta H \sim \frac{mw_\perp^2}{B} \frac{\delta v_\rho}{\rho_{\rm th}} \delta B_{\parallel \rho} \Delta t.
\label{eq:dH2} 
\end{equation} 

Although we are in the process of estimating the rate at which $\mu$
changes over long times, our estimate of $\Delta H$ is comparable to
the value that would follow from $\mu$ conservation:
$\Delta H \sim \mu \Delta B \sim (mw_\perp^2/B) \omega_{\rm eff}
\delta B_{\parallel \rho} \Delta t$,
where
$\Delta B \sim \omega_{\rm eff} \delta B_{\parallel \rho} \Delta t$ is
the rms amplitude of the change in the magnetic flux through the
proton's Larmor orbit, divided by $\pi \rho^2$, during the
time~$\Delta t$ in which the proton is (in our simple two-step model
of proton motion) undergoing continuous, circular, cyclotron
motion. This correspondence highlights an alternative interpretation
of the stochastic-heating process at $\beta_{\rm p} \gtrsim 1$. In the
guiding-center approximation, when $v_\perp^2$ increases by some
factor because
of~$\bm{E}_{\rm s}$, the field strength at the particle's guiding
center increases by approximately the same factor, essentially because of
Faraday's law. This proportionality underlies $\mu$ conservation. In
stochastic heating, the same proportionality is approximately
satisfied during a single time
interval~$\Delta t$, but the proton is then stochastically transported
to a neighboring set of gyroscale eddies, in which the field strength
is not correlated with the field strength at the proton's original
location. The proton thus ``forgets'' about what happened to the field
strength at its original location and gets to keep the energy that it
gained without ``paying the price'' of residing in a higher-field-strength
location. In this way, spatial diffusion perpendicular to~$\bm{B}$
breaks the connection between changes to~$v_\perp^2$ and changes
to~$B$ that arises in the $\rho/\lambda \rightarrow 0$ limit.

In our simple model, the energy gained by a proton is in the form of perpendicular kinetic energy,
\begin{equation}
K_\perp = \frac{m v_\perp^2}{2},
\label{eq:defKperp} 
\end{equation} 
because we neglect the parallel motion of protons. (We do not preclude the possibility of parallel stochastic heating, but we do not consider it further here.)
The perpendicular-kinetic-energy diffusion coefficient $D_{K}$ is thus $\sim \Delta H^2/\Delta t$, or
\begin{equation} 
D_{K}
\sim \frac{m^2 \wperp^4}{\beta_{\rm p}^{1/2}} \frac{ \delta v_{\rho}}{\rho_{\rm th}} \frac{ \delta B_{\rho}^2 }{B_0^2}, 
\label{eq:DK0} 
\end{equation} 
where we have used~(\ref{eq:Dt3}) to estimate $\Delta t$ and~(\ref{eq:eigen})  to set $\delta B_{\parallel \rho} \sim \delta B_\rho$.
A single proton undergoing a random walk in energy can gain or lose energy with equal probability during a time $\Delta t$. However, if a large number of thermal protons (e.g., with an initially Maxwellian distribution)
undergo energy diffusion, then on average more protons will gain energy than lose energy, leading to proton heating.
The heating time scale $\tau_{\rm h}$ is the characteristic time for the perpendicular kinetic energy of a thermal proton to double, $\tau_{\rm h} \sim (m w_\perp^2)^2/D_{K}$, and the perpendicular-heating rate per unit mass is $Q_{\perp} \sim K_{\perp}/(m\tau_{\rm h}) \sim D_{K}/(mK_{\perp})$, or, 
\begin{equation} 
Q_{\perp} \sim \beta_{\rm p}^{1/2}\,\frac{ (\delta v_{\rho})^3}{\rho_{\rm th}} .
\label{eq:Qperp0} 
\end{equation}

To account for the uncertainties introduced by our numerous order-of-magnitude estimates, we multiply the right-hand side of~(\ref{eq:Qperp0}) by an as-yet-unknown dimensionless constant~$\sigma_1$. 
As $\delta v_{\rho} \rightarrow 0$, $\d\mu/\d t$ decreases faster than any power of~$\delta v_\rho$ \citep{kruskal}. To account for this ``exponential'' $\mu$ conservation in the small-$\delta v_\rho$ limit, we follow \cite{cetal2010} by multiplying the right-hand side of~(\ref{eq:Qperp0})  by
the factor $\exp (-\sigma_2/\delta)$, 
\begin{equation} 
Q_{\perp} = \sigma_1\frac{ (\delta v_{\rho})^3}{\rho_{\rm th}} \sqrt{\beta_{\rm p}} \exp \left( -\frac{\sigma_2}{\delta} \right),
\label{eq:Qperp1} 
\end{equation}
where $\sigma_2$ is another as-yet-unknown dimensionless constant, and $\delta$ is 
defined in~(\ref{eq:defdelta}).

For comparison, the stochastic-heating rate per unit mass found by \citet{cetal2010} when  $\beta_{\rm p} \lesssim 1$ is
\begin{equation} 
Q_{\perp} =  c_1 \frac{(\delta v_{\rho})^3}{\rho_{\rm th}}\exp \left( -\frac{c_2}{\varepsilon} \right),
\label{eq:Qperp2} 
\end{equation}
where 
\begin{equation}
\varepsilon = \frac{\delta v_\rho}{\wperp},
\label{eq:eps1}
\end{equation}
and the dimensionless constants $c_1$ and $c_2$ serve the same purpose as those in~(\ref{eq:Qperp1}).
As discussed by \cite{cetal2010} for the case of $c_1$ and~$c_2$, we expect the constants $\sigma_1$  and~$\sigma_2$ to depend on the nature of the fluctuations. For example, at fixed $\delta v_{\rho}$, we expect stronger heating rates (i.e., larger $\sigma_1$ and/or smaller $\sigma_2$) from intermittent turbulence than from randomly phased waves \citep{cetal2010,xetal2013,mallet18}, because, in intermittent turbulence, most of the heating takes place near coherent structures in which the fluctuations are unusually strong and in which the proton orbits are more stochastic than on average. 

\subsection{Orbit stochasticity from parallel motion}
\label{sec:parallel} 

In Section~\ref{sec:perp}, we focused on proton motion perpendicular
to~$\bm{B}$. However, motion along the magnetic field can also produce
stochastic motion in the plane perpendicular to~$\bm{B}_0$ \citep[see, e.g.,][]{hauff10}.
In particular, the perpendicular magnetic fluctuations
at the scale of a proton's gyroradius perturb the direction
of~$\bm{\hat{b}}$. These perturbations, when fed into the first term on the right-hand side of
(\ref{eq:dRdt1}), $v_\parallel \bm{\hat{b}}$, cause the
proton's guiding center~$\bm{R}$ to acquire a velocity perpendicular to~$\bm{B}_0$ of
\begin{equation}
\bm{u}_\perp \sim v_\parallel \times \frac{\delta \bm{B}_{\perp \rho}}{B_0} ,
\label{eq:uperp} 
\end{equation} 
where $\delta \bm{B}_{\perp \rho}$ is the component of~$\delta \bm{B}$
(from gyroscale fluctuations)
perpendicular to~$\bm{B}_0$ at the proton's location. The value of $\bm{u}_\perp$ varies in an incoherent
manner in time, with a correlation time $\sim \Omega^{-1}$. If $u_\perp$ is a significant fraction
of~$v_\perp$, then $\bm{u}_\perp$ will cause a proton's orbit in the
plane perpendicular to~$\bm{B}_0$ to become stochastic.
This leads to an alternative high-$\beta_{\rm p}$ stochasticity parameter,
\begin{equation}
\tilde{\delta} = \frac{u_\perp}{v_\perp} = \frac{v_\parallel \delta B_{\perp
  \rho}}{v_\perp B_0} .
\label{eq:delta2} 
\end{equation} 
As $\tilde{\delta}$ increases towards unity, proton orbits become stochastic.
For thermal
protons with $v_\perp \sim v_\parallel$ and $\rho\sim \rho_{\rm th}$,
$\tilde{\delta}$ is equivalent to $\delta$ in~(\ref{eq:defdelta}), which
was based upon the parallel magnetic-field fluctuation $\delta
B_{\parallel \rho}$ (even though we set $\delta B_{\parallel \rho} \sim
\delta B_\rho$ in~(\ref{eq:defdelta}))). The contribution of parallel motion to orbit
stochasticity thus does not change our conclusions about the rate at
which thermal protons are heated stochastically. However, the
contribution of parallel motion to orbit stochasticity should be taken into account when
considering the ability of stochastic heating to produce superthermal
tails, because in AW turbulence the perpendicular (parallel) magnetic fluctuation at
perpendicular scale~$\lambda$, denoted $\delta B_{\perp \lambda}$
($\delta B_{\parallel \lambda}$), is an increasing (decreasing)
function of~$\lambda$ when $\lambda$ is in the inertial range. Orbit
stochasticity through the interaction between parallel motion and
$\delta B_{\perp \rho}$ could thus contribute to the
development of superthermal tails when $\beta_{\rm p}\gtrsim 1$. An
investigation of superthermal tails, however, lies beyond the scope of this paper.

\section{Numerical Test-Particle Calculations}\label{sec:test}

To test the phenomenological theory developed in Section~\ref{sec:sh}, we numerically track test-particle protons interacting with a spectrum of low-frequency randomly phased AWs and KAWs.
The initial particle positions are random and uniformly distributed within a cubical region of volume $(100d_{\rm p})^3$, where $d_{\rm p} = v_{\rm A}/\Omega$ is the proton inertial length. The initial velocity distribution is an isotropic Maxwellian with proton temperature $T_{\rm p}$. To trace each particle, we solve the equations of motion,
\begin{equation}
\diff{\bx}{t} = \bv
\qquad
\diff{\bv}{t} = \frac{q}{m} \left( \bE + \frac{\bv \times \bB}{c} \right),
\end{equation}
using the Boris method \citep{boris} with a time step of~$0.01 \Omega^{-1}$.

\subsection{Randomly phased waves}\label{sec:spectrum}

The code used to implement the AW/KAW spectrum is similar to the code
used by \citet{cetal2010}. The magnetic field is
$\bB = B_0\zhat + \delta \bB$, where $B_0$ is constant. We take $\bE$
and $\delta\bB$ to be the sum of the electric and magnetic fields of
waves at each of 81 different wave vectors, with two waves of equal
amplitude at each wave vector, one with $\omega/k_z < 0$ and the other
with $\omega/k_z > 0$. \footnote{This makes the cross helicity
  zero. For a discussion of how cross helicity affects the stochastic
  heating rate in the low-$\beta_{\rm p}$ regime, see \cite{chandran13}.}
The initial phase of each wave is randomly chosen.

The 81 wave vectors correspond to nine evenly spaced values of the azimuthal angle in $\bm{k}$ space (in cylindrical coordinates aligned with~$\bm{B}_0$) at each of nine specific values of $k_{\perp i}: i\in {[0,\ldots,8]}$. The values of $k_{\perp i}$ are evenly spaced in $\ln(\kperp)$-space, with $\ln(k_{\perp i}\rho_{\rm th}) = -4/3 + i/3$. The middle three cells, in which $i = 3,4,$ and $5$, have a combined width of unity in $\ln(\kperp)$-space, centred at precisely $\kperp\rho_{\rm th}=1$. We computationally evaluate $\delta v_{\rho}$ and $\delta B_{\rho}$ via the rms values of the $\bE \times \bB$ velocity and $\delta \bB$ that result from the waves in just these middle three cells. 

There is one value of $\kpara \equiv |k_z|$ at each $k_{\perp i}$, denoted $k_{\para i}$. We determine $k_{\para 4}$ by setting the linear frequency at $k_{\para 4}$ equal to $k_{\perp 4} \delta v_{\rho}$. At other values of $\kperp$, we set
\begin{equation} \label{exponents}
\frac{k_{\para i}}{k_{\para 4}} = 
\begin{dcases}
\left(\frac{k_{\perp i}}{k_{\perp 4}}\right)^{2/3} &: 0\leq i<4\\ 
\left(\frac{k_{\perp i}}{k_{\perp 4}}\right)^{1/3} &: 4<i\leq8.
\end{dcases}
\end{equation}
The exponents $2/3$ and $1/3$ in~\eqref{exponents} are chosen to match the scalings in the critical-balance models of \citet{gs95} and \citet{cholaz2004}, respectively.
We take the individual wave magnetic-field amplitudes 
to be proportional to $\kperp^{-1/3}$ and $\kperp^{-2/3}$ for $\kperp\rho_{\rm th}<1$ and $\kperp\rho_{\rm th}>1$, respectively, in order to match the same two critical-balance models. (All the waves at the same value of $k_{\perp i}$ have the same amplitudes.)
We determine the wave frequency and relative amplitudes of the different components of the fluctuating electric and magnetic fields using Hollweg's~(1999) \nocite{hollweg1} two-fluid analysis of linear KAWs, setting
\begin{equation}
\frac{T_{\rm e}}{T_{\rm p}} = 0.5 \qquad \frac{v_{\rm A}}{c} = 0.003,
\label{eq:params} 
\end{equation} 
where $T_{\rm e}$ and $T_{\rm p}$ are the (isotropic) electron and proton temperatures.
We do not expect the particular choices in~(\ref{eq:params}) to have a large effect on our results, but choose those values to facilitate a direct comparison to the previous numerical results of \citet{cetal2010}.
Since we take $T_{\perp \rm p} = T_{\parallel \rm p}$, we set
\begin{equation}
w_\perp = w_\parallel = w \equiv \sqrt{\frac{2k_{\rm B} T_{\rm p}}{m}}
= v_{\rm A} \sqrt{\beta_{\rm p}}.
\label{eq:defw} 
\end{equation}

\subsection{A Note on the Electric Field}

Following \citet{letal2009}, we correct the electric field because the magnetic field (including its fluctuations, i.e., $\bB = B_0\zhat + \delta \bB$)  in the simulation is not orientated along the $z$-axis. The simulation, however, equates the parallel and perpendicular components of the electric field to the parallel and perpendicular components of the wave electric field that would arise if the magnetic field were aligned exactly on the $z$-axis. The result is a numerical addition of perpendicular electric field terms to the parallel electric field, which, in turn, causes non-physical parallel heating. This may be seen in figure~2 of \citet{cetal2010}. To fix this, we replace the sum of the individual wave electric fields described in Section~\ref{sec:spectrum}, which we denote~$\bE_{\rm wave}$, with the modified electric field $\bE = \bE_{\rm wave} + \bm{\hat{b}} (\bm{\hat{z}}  \bcdot \bE_{\rm wave} - \bm{\hat{b}} \bcdot \bE_{\rm wave})$.

\subsection{Perpendicular Heating}

\begin{figure}
\centerline{\includegraphics[width=\textwidth]{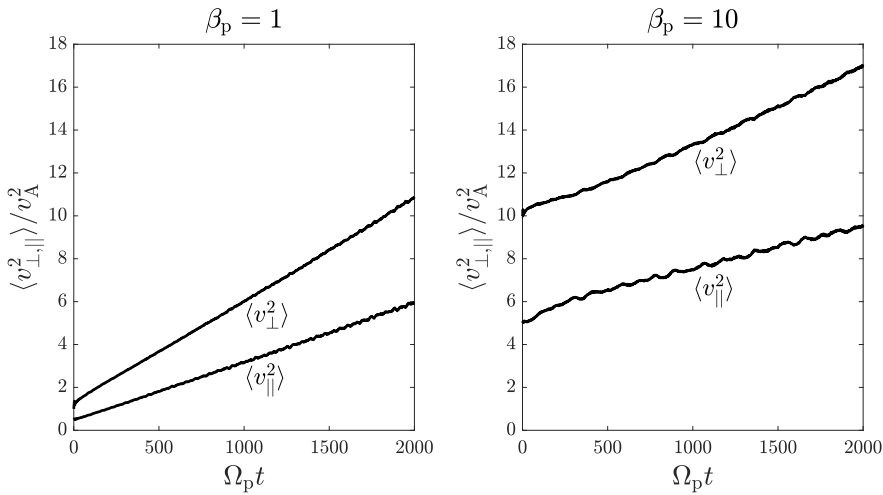}}  
\caption{The mean square velocity perpendicular to $\bm{B}_0$,
  $\langle v_\perp^2 \rangle$, as a function of time in two
  test-particle calculations, each of which tracks $10^5$ protons. The
  value of the stochasticity parameter $\delta$ (defined in
  (\ref{eq:defdelta})) is 0.15 in both calculations. }
\label{fig:Tvst} 
\end{figure}

We perform numerical test-particle calculations at five different values of~$\beta_{\rm p}$,
in particular $\beta_{\rm p} = \{ 0.006, 0.01, 0.1, 1, 10\}$.
For each $\beta_{\rm p}$ value, we carry out a test-particle calculations for five different values of $\delta$ (or, equivalently, five different values of~$\varepsilon$).
Each calculation returns the value of $\langle v_{\perp}^2 \rangle$
and $\langle v_\parallel^2 \rangle$ as
functions of time. We show two examples in figure~\ref{fig:Tvst}.
The slope of the best-fit line for each $\langle\vperp^2(t)\rangle$ curve determines the perpendicular-heating rate per unit mass $Q_{\perp}= (1/2) (\d/\d t)\langle v_{\perp}^2\rangle$, where $\langle \hspace{1.5pt} \cdots \rangle$ indicates an average over the $10^4$ or $10^5$ particles in the simulation. (We use more particles in simulations with smaller $\varepsilon$ and $\delta$ because the heating rates are smaller in these simulations, and the extra particles increase the signal-to-noise ratio.)
We fit the $\langle v_\perp^2(t)\rangle$ curves during the time interval $(t_{\rm i}, t_{\rm f})$, where $t_{\rm i} = 20\pi/\Omega$ and $t_{\rm f}$ is the smaller of the following two values: $10^4 \Omega^{-1}$
and the time required for $\langle v_\perp^2\rangle$ to increase by $\simeq 30\%$. 
We do not include the first ten cyclotron periods when calculating
$Q_{\perp}$, because it takes the particles a few cyclotron periods to
adjust to the presence of the waves, during which time there is
typically strong transient heating. 
(As figure~\ref{fig:Tvst}
  shows, the test particles undergo parallel heating as well as perpendicular
  heating, as was found previously by \cite{xetal2013} in simulations
  of test particles interacting with reduced magnetohydrodynamic turbulence at
  $\beta_{\parallel \rm p} = 1$.)

\begin{figure}
\centerline{\includegraphics[width=\textwidth]{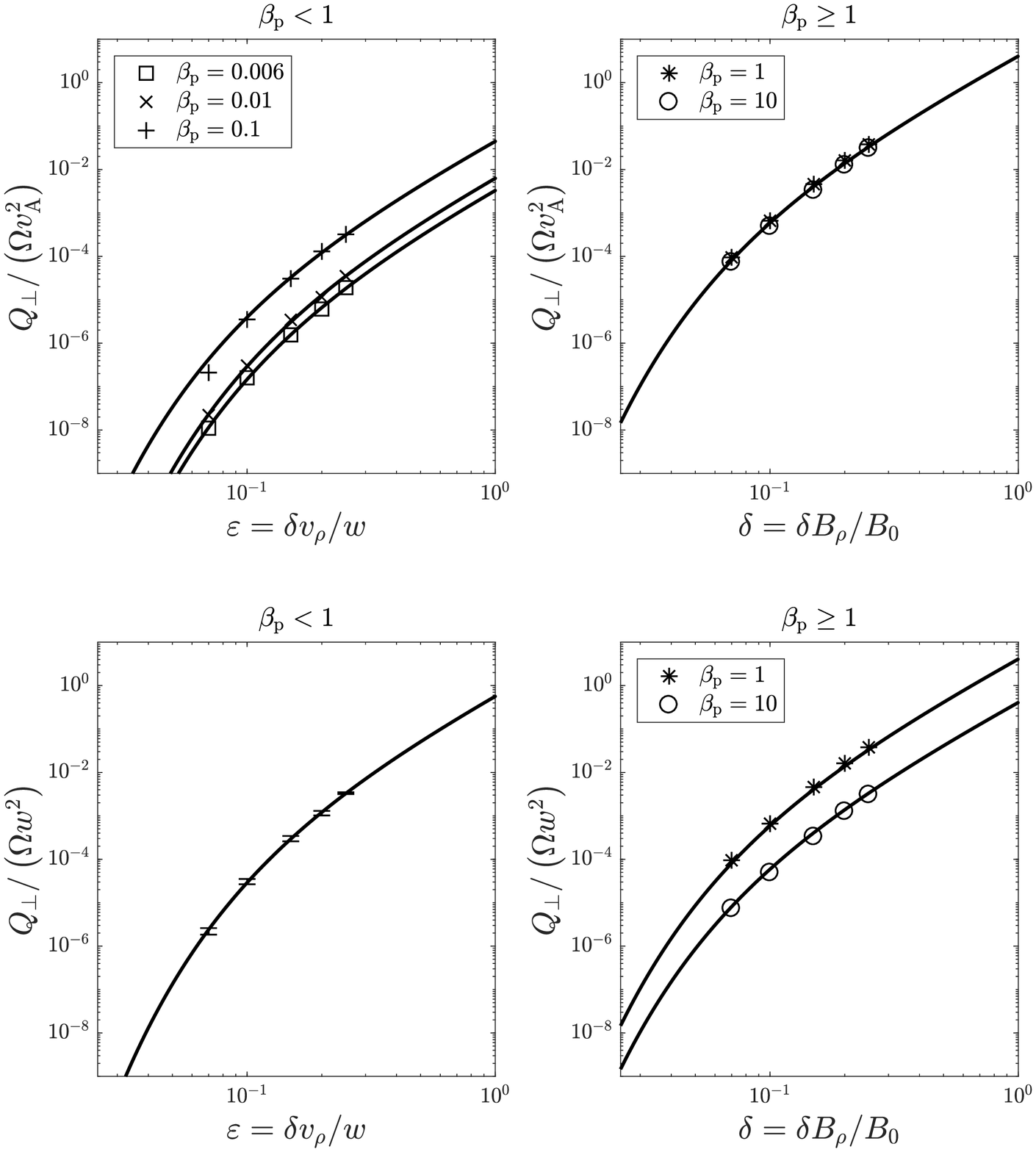}}  
\caption{
Numerical results for the perpendicular-heating rate per unit mass,~$Q_\perp$, for protons interacting with randomly phased AWs and KAWs.
Top-Left: $\beta_{\rm p}<1$, and $Q_\perp$ normalized by $\Omega v_{\rm A}^2$. Top right: $\beta_{\rm p} \geq 1$ and $Q_{\perp}$ normalized by $\Omega v_{\rm A}^2$. Bottom left: $\beta_{\rm p} < 1$ and $Q_\perp$ normalized by $\Omega w^2$, where $w$ is the proton thermal speed
defined in~(\ref{eq:defw});
 the numerical results for all three $\beta_{\rm p}$ values (0.006,
 0.01, and 0.1) are within the bars shown.
 Bottom right: $\beta_{\rm p}\geq 1$ and $Q_\perp$ normalized by $\Omega w^2$.
In the left two panels, the solid lines are plots of
    (\ref{eq:Qperp2}) for the best-fit values of~$c_1$ and $c_2$ given
    in (\ref{eq:c1c2}). In the right two panels, the solid lines are
    plots of (\ref{eq:Qperp1}) for the best-fit values of $\sigma_1$ and
    $\sigma_2$ in (\ref{eq:s1s2}).}
\label{perp}
\end{figure}

The perpendicular-heating rates in our test-particle calculations are
shown in figure~\ref{perp}. The solid-line curves in the two panels on
the left correspond to~(\ref{eq:Qperp2}), with
\begin{equation}
c_1 = 0.77 \qquad c_2 = 0.33.
\label{eq:c1c2} 
\end{equation}
 These values are very similar to the values $c_1 = 0.75$ and $c_2 =
 0.34$ obtained by \cite{cetal2010} at $\beta_{\rm p} = 0.006$. The
 solid-line curves in the two panels on the right correspond
 to~(\ref{eq:Qperp1})  with 
\begin{equation}
\sigma_1 = 5.0 \qquad \sigma_2 = 0.21.
\label{eq:s1s2} 
\end{equation} 
The agreement between our numerical results and~(\ref{eq:Qperp1}) suggests that the approximations used to derive this equation are reasonable.

The lower-left panel of figure~\ref{perp}  shows that at $\beta_{\rm p}<1$, $Q_\perp/(\Omega w^2)$ is a function of~$\varepsilon$ alone, consistent with the fact that~(\ref{eq:Qperp2}) can be rewritten in the form
\begin{equation}
\frac{Q_\perp}{\Omega w^2} = c_1 \varepsilon^3 \exp\left( -\frac{c_2}{\varepsilon}\right)
\hspace{0.3cm} \mbox{ (at $\beta_{\rm p} \lesssim 1$)}.
\label{eq:Qperp3} 
\end{equation} 
The top-right panel of figure~\ref{perp}  shows that at $\beta_{\rm p}\geq1$, 
$Q_\perp/(\Omega v_{\rm A}^2)$ is a function of~$\delta$ alone, consistent with the fact that~(\ref{eq:Qperp1}) can be rewritten as
\begin{equation}
\frac{Q_\perp}{\Omega v_{\rm A}^2} = \sigma_1 \delta^3 \exp\left( -\frac{\sigma_2}{\delta}\right)
\hspace{0.3cm} \mbox{ (at $\beta_{\rm p} \gtrsim 1$)}.
\label{eq:Qperp4} 
\end{equation} 

We note that in our model of randomly phased KAWs \citep{cetal2010},
\begin{equation}
\frac{\delta B_\rho}{B_0} = 0.84 \frac{\delta v_{\rho}}{v_{\rm A}} ,
\label{eq:dbdv} 
\end{equation} 
and thus
\begin{equation}
\delta = \frac{\delta B_\rho}{B_0}  = 0.84 \frac{\delta v_{\rho}}{v_{\rm A}} 
= 0.84 \beta_{\rm p}^{1/2} \frac{\delta v_\rho}{\wperp} = 0.84 \beta_{\rm p}^{1/2} \varepsilon.
\label{eq:eps1eps2} 
\end{equation} 
As a consequence, if we adopt the best-fit values of $\sigma_1$, $\sigma_2$, $c_1$, and~$c_2$, then
the value of $Q_\perp$ at $\beta_{\rm p} = 1$ in (\ref{eq:Qperp1}),
which matches our test-particle calculations quite well, exceeds the
value that would follow from (\ref{eq:Qperp2}) at $\beta_{\rm p}=1$.
A similar phenomenon was found by \cite{xetal2013} in numerical simulations of test particles interacting with strong reduced magnetohydrodynamic turbulence.

\begin{figure}
\centerline{\includegraphics[width=0.5\textwidth]{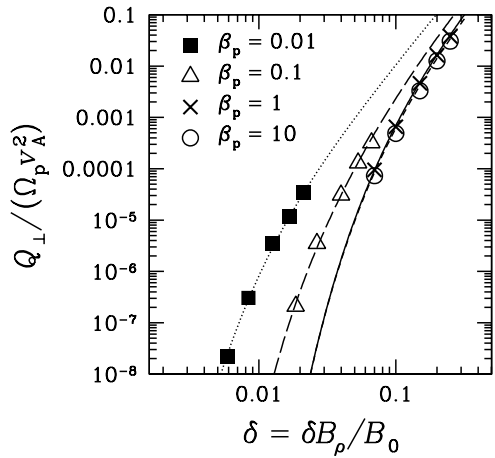}}  
\caption{The data points reproduce the numerical results from
  figure~\ref{perp} for
  $\beta_{\rm p} = 0.01, 0.1, 1.0, \mbox{ and } 10$. The dotted,
  long-dashed, solid, and short-dashed lines plot
  (\ref{eq:unified}) for, respectively, $\beta_{\rm p} =0.01$,
  $\beta_{\rm p} = 0.1$, $\beta_{\rm p} = 1$, and
  $\beta_{\rm p} = 10$. The solid and short-dashed lines are difficult
  to distinguish because they are nearly on top of each other.}
\label{merged}
\end{figure}

To obtain a fitting formula that can be used to model stochastic
  heating at large~$\beta_{\rm p}$, small~$\beta_{\rm p}$, and
  $\beta_{\rm p} \simeq 1$, we use (\ref{eq:defw}) and (\ref{eq:eps1eps2})
  to rewrite the low-$\beta_{\rm p}$ heating rate in (\ref{eq:Qperp3})
  in terms of $\delta$ and $v_{\rm A}$. We then add the
  low-$\beta_{\rm p}$ heating rate to the high-$\beta_{\rm p}$ heating
  rate in (\ref{eq:Qperp4}), obtaining
\begin{equation}
\frac{Q_\perp}{\Omega v_{\rm A}^2} = \sigma_1 \delta^3 \exp\left( -\frac{\sigma_2}{\delta}\right)
+ \frac{1.69 c_1 \delta^3}{\beta_{\rm p}^{1/2}} \exp\left(- \frac{0.84
    c_2 \beta_{\rm p}^{1/2}}{\delta}\right).
\label{eq:unified} 
\end{equation} 
The first term on the right-hand side dominates at
$\beta_{\rm p} \gtrsim 1$ in part because $\sigma_1 \simeq 6.5 c_1$.
The second term on the right-hand side dominates at
$\beta_{\rm p} \ll 1$.  Figure~\ref{merged} shows that
(\ref{eq:unified}) is consistent with our numerical results. This
figure also illustrates how, at fixed $\delta B_{\rho}/B_0$, the
stochastic heating rate increases as $\beta_{\rm p}$ decreases.  

As mentioned above, stochastic heating becomes more effective as the
fluctuations become more intermittent \citep{xetal2013,mallet18}.
The randomly phased waves in our test-particle simulations are not
intermittent, but gyroscale fluctuations in space and astrophysical
plasmas generally are \citep[see, e.g.,][]{mangeney01,carbone04,salem09,chandran15,mallet15}.
Further work is needed to determine how the best-fit constants in
\eqref{eq:c1c2} and \eqref{eq:s1s2} depend upon the degree of
intermittency at the proton gyroradius scale. Until this dependency is
determined, some caution should be exercised when applying \eqref{eq:unified} to space and
astrophysical plasmas. For reference, \cite{bourouaine13a} found that 
lowering $c_2$ to $\simeq 0.2$ led the
heating rate in (\ref{eq:Qperp2}) to be consistent with the proton
heating rate and fluctuation amplitudes inferred from measurements of
the fast solar wind from the {\em Helios} spacecraft at
$r=0.3 \mbox{ au}$.  However, if $c_2 = 0.33$, then the heating rate
in (\ref{eq:Qperp2}) is too weak to explain the proton heating seen in
the {\em Helios} measurements.

\section{Summary}\label{sec:conclusion}

In this paper we use phenomenological arguments to derive an analytic
formula for the rate at which thermal protons are stochastically
heated by AW/KAW turbulence at $k_\perp \rho_{\rm th} \sim 1$. We
focus on $\beta_{\rm p} \sim 1 -30$. Smaller values of~$\beta_{\rm p}$
were considered by \citet{cetal2010}. At larger values
of~$\beta_{\rm p}$, KAWs at $k_\perp \rho_{\rm th} \sim 1$ become
non-propagating, and some of the scalings we have assumed do not
apply.  At $\beta_{\rm p} \sim 1- 30$, the motion of a proton's
effective guiding center is dominated by the interaction between the
proton and gyroscale fluctuations in the magnetic field,
whose amplitude is denoted~$\delta B_{\rho}$.  As
$\delta B_{ \rho}/B_0$ increases from infinitesimal values
towards unity, the proton motion in the plane perpendicular
to~$\bm{B}_0$ becomes random (stochastic), leading to spatial
diffusion, and this spatial diffusion breaks the strong correlation
between changes in a proton's perpendicular kinetic energy and the
magnetic-field strength at the proton's location that normally gives
rise to magnetic-moment conservation. The interaction between the
proton and the electric field then becomes a Markov process that causes
the proton to diffuse in energy. This energy diffusion leads to
heating. At $\beta_{\rm p} \sim 1-30$, it is the solenoidal component
of the electric field that dominates the heating.

The analytic formula that we derive for the stochastic heating
rate~$Q_\perp$ contains two dimensionless constants, $\sigma_1$ and
$\sigma_2$, whose values depend upon the nature of the AW/KAW
fluctuations that the proton interacts with (e.g., randomly phased
waves or intermittent turbulence). We numerically track test particles
interacting with randomly phased AWs and KAWs and find that our
analytic formula for $Q_\perp$ agrees well with the heating rate of
these test particles for the choices $\sigma_1 = 5.0$ and $\sigma_2 =
0.21$.  We note that previous work has shown that for fixed rms
amplitudes of the gyroscale fluctuations, stochastic heating is more
effective when protons interact with intermittent turbulence than when
protons interact with randomly phased waves
\citep{cetal2010,xetal2013,mallet18}. The reason for this is that in
intermittent turbulence, most of the heating occurs near coherent
structures, in which the fluctuation amplitudes are larger than average and in which the particle orbits are more stochastic than on average.

Our work leaves a number of interesting questions unanswered. Two such
questions are how the energy-diffusion coefficient depends on energy
at $\beta_{\rm p} \sim 1-30$ and how the proton distribution function
evolves in the presence of stochastic heating. (For a discussion of
the low-$\beta_{\rm p}$ case, see \cite{klein16}.)  We also have not
addressed the question of how stochastic heating changes as
$\beta_{\rm p}$ is increased to values $\gtrsim 30$ and KAWs at
$k_\perp \rho_{\rm th} \sim 1$ become non-propagating, or how the
stochastic heating rate for minor ions depends upon minor-ion mass,
charge, and average flow speed along~$\bm{B}_0$ in the proton
frame. (For a discussion of the low-$\beta_{\rm p}$ case, see
\cite{chandran13}.) Previous studies have compared 
  observationally inferred heating rates in the solar wind with the
  low-$\beta_{\parallel \rm p}$ stochastic-heating rate in
  (\ref{eq:Qperp2}) derived by \cite{cetal2010}, finding quantitative
  agreement at $r=0.3 \mbox{ au}$ assuming $c_2 \simeq 0.2$
  \citep{bourouaine13a} and qualitative agreement at $r=1 \mbox{ au}$
  \citep{vech17}. However, it is not yet clear whether the stochastic
  heating rate in (\ref{eq:Qperp1}) agrees with solar-wind measurements
  in the large-$\beta_{\parallel \rm p}$ regime.  In addition,
  stochastic heating at $\beta_{\parallel \rm p} \gtrsim 1$ could
  trigger temperature-anisotropy instabilities, which could in turn
  modify the rate(s) of perpendicular (and parallel) proton heating.
Future investigations of these questions will be important for
determining more accurately the role of stochastic heating in space
and astrophysical plasmas.

We thank L. Arzamasskiy, Y. Kawazura, M. Kunz, E. Quataert, and A. Schekochihin for valuable discussions.
This work was supported in part by NASA grants NNX15AI80, NNX16AG81G,
NNS16AM23G, NNX17AI18G, and NNN06AA01C, and  NSF grant PHY-1500041. D.~Verscharen acknowledges the support of
STFC Ernest Rutherford Fellowship ST/P003826/1.

\appendix

\section{Non-propagation of KAWs at $k_\perp \rho_{\rm th} \sim 1$ at high~$\beta_{\rm p}$}
\label{sec:limitations} 

\begin{figure}
\centerline{\includegraphics[width=\textwidth]{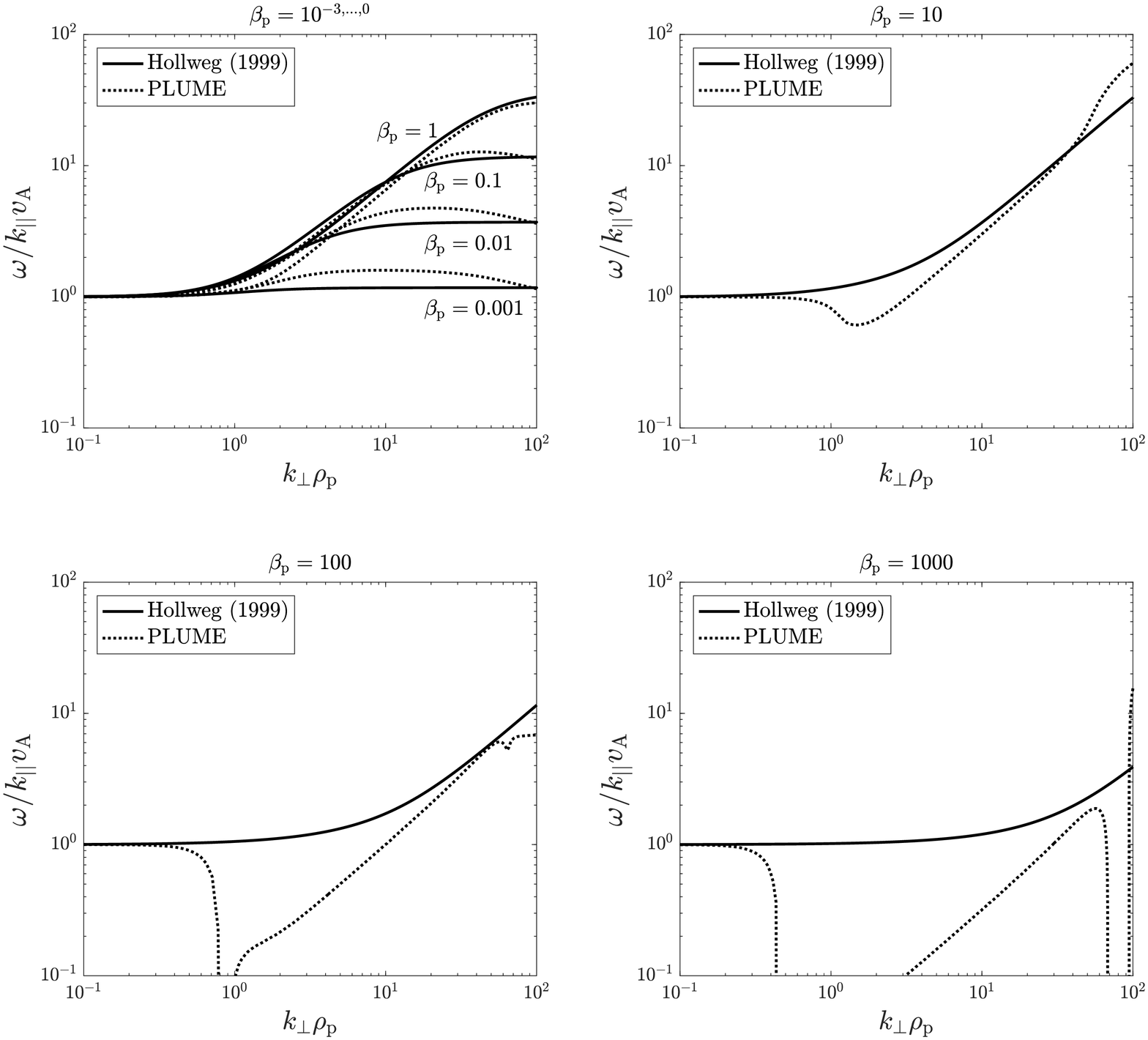}}
\caption{The KAW dispersion relation from Hollweg's (1999) two-fluid model (solid lines) and the PLUME hot-plasma dispersion-relation solver \citep{plume} (dotted lines) for $T_{\rm e} /T_{\rm p} = 0.5$ and
$v_{\rm A}/c = 0.003$ for various powers of~$\beta_{\rm p}$.
\label{plume}}
\end{figure}

In figure \ref{plume}, we compare the AW/KAW dispersion relation from the two-fluid model of \citet{hollweg1} and the PLUME hot-plasma dispersion-relation solver \citep{plume} for $T_{\rm e}/T_{\rm p} = 0.5$ and $v_{\rm A}/c = 0.003$ and for various values of $\beta_{\rm p}$.  The PLUME results shown here assume 
that $k_\parallel \rho_{\rm th} = 0.001$ and that the proton and electron distributions are
Maxwellian. The two-fluid dispersion relation agrees reasonably well with the more accurate PLUME results at $\beta_{\rm p} \lesssim 1$. However, at $\beta_{\rm p} \gtrsim 1$, the PLUME results deviate from the two-fluid theory because of ion damping, which becomes stronger as $\beta_{\rm p}$ increases \citep{2howes2008,kunz18}.  Starting at $\beta_{\rm p}\simeq 30$ (for  $T_{\rm e}/T_{\rm p} = 0.5$, $v_{\rm A}/c = 0.003$, and $k_\parallel \rho_{\rm p} = 0.001$), the real part (but not the imaginary part) of the KAW frequency at $k_\perp \rho_{\rm th} = 1$ vanishes (i.e., KAWs become damped, non-propagating modes). For larger $\beta_{\rm p}$ values, KAWs are non-propagating throughout an interval of $k_\perp \rho_{\rm th}$ values centered on unity that broadens to both larger and smaller values as~$\beta_{\rm p}$ increases \citep{kawazura18}.

\section{Stochastic heating by the electrostatic potential at $\beta_{\rm p}\gtrsim 1$}
\label{ap:potential} 

In Section~\ref{sec:sh} we considered the rms change to a thermal proton's energy $\Delta H$ resulting from the solenoidal component of the electric field $\bm{E}_{\rm s}$ during the particle residence time~$\Delta t$ within one set of gyroscale eddies. We also evaluated the contribution of $\bm{E}_{\rm s}$ to the stochastic heating rate~$Q_{\perp}$. Here, we show that the contribution to~$Q_\perp$ from~$\bm{E}_{\rm s}$ is larger than the contribution from the electrostatic potential~$\Phi$ when $\beta_{\rm p} \gtrsim 1$.

We assume that the rms amplitude of the potential part of the electric field at $k_\perp \rho_{\rm th}$ is comparable to the rms amplitude of the total gyroscale electric-field fluctuation,~$\delta E_{\rho}$, which in turn is~$\sim \delta v_\rho B_0/c$.
As discussed by \cite{cetal2010}, the contribution of the time-varying
electrostatic potential to~$\Delta H$ is
\begin{equation}
\Delta H_{\rm potential} \sim q \omega_{\rm eff} \Delta \Phi_{\rho} \Delta t,
\label{eq:dHA} 
\end{equation}
where $\omega_{\rm eff} = \delta v_\rho/\rho_{\rm th}$ (see~(\ref{eq:omegaeff})), and
\begin{equation}
q \Delta\Phi_{\rho} \sim q \rho_{\rm th} \delta E_{\rho} \sim q w_\perp \times \frac{mc}{qB_0 }\times \frac{\delta v_\rho B_0}{c} \sim m w_\perp \delta v_\rho .
\label{eq:qdPhi} 
\end{equation} 
Since~(\ref{eq:Dt3}) gives $\Delta t \sim \beta_{\rm p}^{-1/2} \rho_{\rm th}/\delta v_\rho$,
\begin{equation}
\omega_{\rm eff} \Delta t \sim \beta_{\rm p}^{-1/2}.
\label{eq:omdt} 
\end{equation} 
Combining~(\ref{eq:dHA}) through (\ref{eq:omdt}), we obtain
\begin{equation}
\Delta H_{\rm potential} \sim \beta_{\rm p}^{-1/2} m w_\perp \delta v_\rho,
\label{eq:dHA2} 
\end{equation} 
\begin{equation}
D_{K, \rm potential} \sim \frac{(\Delta H_{\rm potential})^2}{\Delta t} \sim \frac{\beta_{\rm p}^{-1} m^2 w_\perp^2 (\delta v_\rho)^2}{\beta_{\rm p}^{-1/2} \rho_{\rm th}/\delta v_\rho} \sim \beta_{\rm p}^{-1/2} m^2 w_\perp^2 \frac{(\delta v_{\rho})^{3}}{\rho_{\rm th}},
\label{eq:DKA} 
\end{equation} 
and
\begin{equation}
Q_{\perp, \rm potential} \sim \frac{D_{K, \rm potential}}{m K_\perp} \sim \beta_{\rm p}^{-1/2} 
\frac{(\delta v_{\rho})^{3}}{\rho_{\rm th}},
\label{eq:QperpA} 
\end{equation} 
which is a factor of~$\sim \beta_{\rm p}^{-1}$ smaller than the estimate of~$Q_\perp$ in~(\ref{eq:Qperp0}).

\bibliographystyle{jpp}

\bibliography{jpp-instructions}

\end{document}